\documentclass[twocolumn,letterpaper,aps,prd,superscriptaddress,nofootinbib,longbibliography,floatfix]{revtex4-1}
\usepackage{graphicx}	
\usepackage{xspace}     

\begin{document}


\title{Cross section and longitudinal single-spin asymmetry $A_L$ for 
forward $W^{\pm}\rightarrow\mu^{\pm}\nu$ production in polarized $p$$+$$p$ 
collisions at $\sqrt{s}=510$ GeV. }

\newcommand{\abilene}{Abilene Christian University, Abilene, Texas 79699, USA}
\newcommand{\augie}{Department of Physics, Augustana University, Sioux Falls, South Dakota 57197, USA}
\newcommand{\banaras}{Department of Physics, Banaras Hindu University, Varanasi 221005, India}
\newcommand{\barc}{Bhabha Atomic Research Centre, Bombay 400 085, India}
\newcommand{\baruch}{Baruch College, City University of New York, New York, New York, 10010 USA}
\newcommand{\bnlcoll}{Collider-Accelerator Department, Brookhaven National Laboratory, Upton, New York 11973-5000, USA}
\newcommand{\bnlphys}{Physics Department, Brookhaven National Laboratory, Upton, New York 11973-5000, USA}
\newcommand{\caucr}{University of California-Riverside, Riverside, California 92521, USA}
\newcommand{\charlesczech}{Charles University, Ovocn\'{y} trh 5, Praha 1, 116 36, Prague, Czech Republic}
\newcommand{\chonbuk}{Chonbuk National University, Jeonju, 561-756, Korea}
\newcommand{\ciae}{Science and Technology on Nuclear Data Laboratory, China Institute of Atomic Energy, Beijing 102413, People's Republic of China}
\newcommand{\cns}{Center for Nuclear Study, Graduate School of Science, University of Tokyo, 7-3-1 Hongo, Bunkyo, Tokyo 113-0033, Japan}
\newcommand{\colorado}{University of Colorado, Boulder, Colorado 80309, USA}
\newcommand{\columbia}{Columbia University, New York, New York 10027 and Nevis Laboratories, Irvington, New York 10533, USA}
\newcommand{\czechtech}{Czech Technical University, Zikova 4, 166 36 Prague 6, Czech Republic}
\newcommand{\debrecen}{Debrecen University, H-4010 Debrecen, Egyetem t{\'e}r 1, Hungary}
\newcommand{\elte}{ELTE, E{\"o}tv{\"o}s Lor{\'a}nd University, H-1117 Budapest, P{\'a}zm{\'a}ny P.~s.~1/A, Hungary}
\newcommand{\eszterhazy}{Eszterh\'azy K\'aroly University, K\'aroly R\'obert Campus, H-3200 Gy\"ongy\"os, M\'atrai \'ut 36, Hungary}
\newcommand{\ewha}{Ewha Womans University, Seoul 120-750, Korea}
\newcommand{\fsu}{Florida State University, Tallahassee, Florida 32306, USA}
\newcommand{\gsu}{Georgia State University, Atlanta, Georgia 30303, USA}
\newcommand{\hanyang}{Hanyang University, Seoul 133-792, Korea}
\newcommand{\hiroshima}{Hiroshima University, Kagamiyama, Higashi-Hiroshima 739-8526, Japan}
\newcommand{\howard}{Department of Physics and Astronomy, Howard University, Washington, DC 20059, USA}
\newcommand{\ihepprot}{IHEP Protvino, State Research Center of Russian Federation, Institute for High Energy Physics, Protvino, 142281, Russia}
\newcommand{\illuiuc}{University of Illinois at Urbana-Champaign, Urbana, Illinois 61801, USA}
\newcommand{\inrras}{Institute for Nuclear Research of the Russian Academy of Sciences, prospekt 60-letiya Oktyabrya 7a, Moscow 117312, Russia}
\newcommand{\instpasczech}{Institute of Physics, Academy of Sciences of the Czech Republic, Na Slovance 2, 182 21 Prague 8, Czech Republic}
\newcommand{\isu}{Iowa State University, Ames, Iowa 50011, USA}
\newcommand{\jaea}{Advanced Science Research Center, Japan Atomic Energy Agency, 2-4 Shirakata Shirane, Tokai-mura, Naka-gun, Ibaraki-ken 319-1195, Japan}
\newcommand{\jyvaskyla}{Helsinki Institute of Physics and University of Jyv{\"a}skyl{\"a}, P.O.Box 35, FI-40014 Jyv{\"a}skyl{\"a}, Finland}
\newcommand{\kek}{KEK, High Energy Accelerator Research Organization, Tsukuba, Ibaraki 305-0801, Japan}
\newcommand{\korea}{Korea University, Seoul, 136-701, Korea}
\newcommand{\kurchatov}{National Research Center ``Kurchatov Institute", Moscow, 123098 Russia}
\newcommand{\kyoto}{Kyoto University, Kyoto 606-8502, Japan}
\newcommand{\labllr}{Laboratoire Leprince-Ringuet, Ecole Polytechnique, CNRS-IN2P3, Route de Saclay, F-91128, Palaiseau, France}
\newcommand{\lahorelums}{Physics Department, Lahore University of Management Sciences, Lahore 54792, Pakistan}
\newcommand{\lawllnl}{Lawrence Livermore National Laboratory, Livermore, California 94550, USA}
\newcommand{\losalamos}{Los Alamos National Laboratory, Los Alamos, New Mexico 87545, USA}
\newcommand{\lund}{Department of Physics, Lund University, Box 118, SE-221 00 Lund, Sweden}
\newcommand{\lyon}{IPNL, CNRS/IN2P3, Univ Lyon, Université Lyon 1, F-69622, Villeurbanne, France}
\newcommand{\maryland}{University of Maryland, College Park, Maryland 20742, USA}
\newcommand{\mass}{Department of Physics, University of Massachusetts, Amherst, Massachusetts 01003-9337, USA}
\newcommand{\michigan}{Department of Physics, University of Michigan, Ann Arbor, Michigan 48109-1040, USA}
\newcommand{\muhlenberg}{Muhlenberg College, Allentown, Pennsylvania 18104-5586, USA}
\newcommand{\myongji}{Myongji University, Yongin, Kyonggido 449-728, Korea}
\newcommand{\nagasaki}{Nagasaki Institute of Applied Science, Nagasaki-shi, Nagasaki 851-0193, Japan}
\newcommand{\nara}{Nara Women's University, Kita-uoya Nishi-machi Nara 630-8506, Japan}
\newcommand{\natmephi}{National Research Nuclear University, MEPhI, Moscow Engineering Physics Institute, Moscow, 115409, Russia}
\newcommand{\newmex}{University of New Mexico, Albuquerque, New Mexico 87131, USA}
\newcommand{\nmsu}{New Mexico State University, Las Cruces, New Mexico 88003, USA}
\newcommand{\ohio}{Department of Physics and Astronomy, Ohio University, Athens, Ohio 45701, USA}
\newcommand{\ornl}{Oak Ridge National Laboratory, Oak Ridge, Tennessee 37831, USA}
\newcommand{\orsay}{IPN-Orsay, Univ.~Paris-Sud, CNRS/IN2P3, Universit\'e Paris-Saclay, BP1, F-91406, Orsay, France}
\newcommand{\peking}{Peking University, Beijing 100871, People's Republic of China}
\newcommand{\pnpi}{PNPI, Petersburg Nuclear Physics Institute, Gatchina, Leningrad region, 188300, Russia}
\newcommand{\riken}{RIKEN Nishina Center for Accelerator-Based Science, Wako, Saitama 351-0198, Japan}
\newcommand{\rikjrbrc}{RIKEN BNL Research Center, Brookhaven National Laboratory, Upton, New York 11973-5000, USA}
\newcommand{\rikkyo}{Physics Department, Rikkyo University, 3-34-1 Nishi-Ikebukuro, Toshima, Tokyo 171-8501, Japan}
\newcommand{\saispbstu}{Saint Petersburg State Polytechnic University, St.~Petersburg, 195251 Russia}
\newcommand{\seoulnat}{Department of Physics and Astronomy, Seoul National University, Seoul 151-742, Korea}
\newcommand{\stonybrkc}{Chemistry Department, Stony Brook University, SUNY, Stony Brook, New York 11794-3400, USA}
\newcommand{\stonycrkp}{Department of Physics and Astronomy, Stony Brook University, SUNY, Stony Brook, New York 11794-3800, USA}
\newcommand{\sungskku}{Sungkyunkwan University, Suwon, 440-746, Korea}
\newcommand{\tenn}{University of Tennessee, Knoxville, Tennessee 37996, USA}
\newcommand{\titech}{Department of Physics, Tokyo Institute of Technology, Oh-okayama, Meguro, Tokyo 152-8551, Japan}
\newcommand{\tsukuba}{Tomonaga Center for the History of the Universe, University of Tsukuba, Tsukuba, Ibaraki 305, Japan}
\newcommand{\vandy}{Vanderbilt University, Nashville, Tennessee 37235, USA}
\newcommand{\weizmann}{Weizmann Institute, Rehovot 76100, Israel}
\newcommand{\wigner}{Institute for Particle and Nuclear Physics, Wigner Research Centre for Physics, Hungarian Academy of Sciences (Wigner RCP, RMKI) H-1525 Budapest 114, POBox 49, Budapest, Hungary}
\newcommand{\yonsei}{Yonsei University, IPAP, Seoul 120-749, Korea}
\newcommand{\zagreb}{Department of Physics, Faculty of Science, University of Zagreb, Bijeni\v{c}ka c.~32 HR-10002 Zagreb, Croatia}
\affiliation{\abilene}
\affiliation{\augie}
\affiliation{\banaras}
\affiliation{\barc}
\affiliation{\baruch}
\affiliation{\bnlcoll}
\affiliation{\bnlphys}
\affiliation{\caucr}
\affiliation{\charlesczech}
\affiliation{\chonbuk}
\affiliation{\ciae}
\affiliation{\cns}
\affiliation{\colorado}
\affiliation{\columbia}
\affiliation{\czechtech}
\affiliation{\debrecen}
\affiliation{\elte}
\affiliation{\eszterhazy}
\affiliation{\ewha}
\affiliation{\fsu}
\affiliation{\gsu}
\affiliation{\hanyang}
\affiliation{\hiroshima}
\affiliation{\howard}
\affiliation{\ihepprot}
\affiliation{\illuiuc}
\affiliation{\inrras}
\affiliation{\instpasczech}
\affiliation{\isu}
\affiliation{\jaea}
\affiliation{\jyvaskyla}
\affiliation{\kek}
\affiliation{\korea}
\affiliation{\kurchatov}
\affiliation{\kyoto}
\affiliation{\labllr}
\affiliation{\lahorelums}
\affiliation{\lawllnl}
\affiliation{\losalamos}
\affiliation{\lund}
\affiliation{\lyon}
\affiliation{\maryland}
\affiliation{\mass}
\affiliation{\michigan}
\affiliation{\muhlenberg}
\affiliation{\myongji}
\affiliation{\nagasaki}
\affiliation{\nara}
\affiliation{\natmephi}
\affiliation{\newmex}
\affiliation{\nmsu}
\affiliation{\ohio}
\affiliation{\ornl}
\affiliation{\orsay}
\affiliation{\peking}
\affiliation{\pnpi}
\affiliation{\riken}
\affiliation{\rikjrbrc}
\affiliation{\rikkyo}
\affiliation{\saispbstu}
\affiliation{\seoulnat}
\affiliation{\stonybrkc}
\affiliation{\stonycrkp}
\affiliation{\sungskku}
\affiliation{\tenn}
\affiliation{\titech}
\affiliation{\tsukuba}
\affiliation{\vandy}
\affiliation{\weizmann}
\affiliation{\wigner}
\affiliation{\yonsei}
\affiliation{\zagreb}
\author{A.~Adare} \affiliation{\colorado} 
\author{C.~Aidala} \affiliation{\losalamos} \affiliation{\michigan} 
\author{N.N.~Ajitanand} \altaffiliation{Deceased} \affiliation{\stonybrkc} 
\author{Y.~Akiba} \email[PHENIX Spokesperson: ]{akiba@rcf.rhic.bnl.gov} \affiliation{\riken} \affiliation{\rikjrbrc} 
\author{R.~Akimoto} \affiliation{\cns} 
\author{J.~Alexander} \affiliation{\stonybrkc} 
\author{M.~Alfred} \affiliation{\howard} 
\author{K.~Aoki} \affiliation{\kek} \affiliation{\riken} 
\author{N.~Apadula} \affiliation{\isu} \affiliation{\stonycrkp} 
\author{Y.~Aramaki} \affiliation{\riken} 
% ...
\collaboration{PHENIX Collaboration}  \noaffiliation


\date{\today}

\begin{abstract}


We have measured the cross section and single-spin asymmetries from 
forward $W^{\pm}\rightarrow\mu^{\pm}\nu$ production in longitudinally 
polarized $p$$+$$p$ collisions at $\sqrt{s}=510$~GeV using the PHENIX 
detector at the Relativistic Heavy Ion Collider. The cross sections are 
consistent with previous measurements at this collision energy, while 
the most forward and backward longitudinal single spin asymmetries 
provide new insights into the sea quark helicities in the proton. The 
charge of the W bosons provides a natural flavor separation of the 
participating partons.

\end{abstract}

\maketitle


\section{Introduction}

The spin of the proton and its decomposition is fundamentally important. 
Understanding its origin is essential to explaining how the strong 
interaction, described by quantum chromodynamics QCD, creates the basic 
building blocks of the visible matter in our universe, protons and 
neutrons. Mostly from deep inelastic scattering measurements and 
hadron-hadron collisions, it is known that quarks and gluons make 
roughly equal contributions to the total momentum of the proton in the 
Bjorken frame \cite{Ball:2017nwa,Schmidt:2015zda,Harland-Lang:2014zoa}. 
Just like gluons, sea quarks also play a substantial role in the 
composition of the proton momentum. Unlike what is naively expected from 
gluon splitting, the unpolarized light quark sea is found to be 
asymmetric with more anti-down quarks than anti-up quarks at small to 
intermediate Bjorken $x<$ 0.2, where $x$ is the parton momentum fraction 
in the infinite momentum frame. See, for example, a review of the world 
data on the unpolarized light sea and the theoretical models related to 
it \cite{Chang:2014jba}.

While several models can describe correctly the measured unpolarized 
light sea, these models differ significantly in their predictions for 
the polarized case \cite{Chang:2014jba}. Valence quark helicity 
contributions to the total spin of the nucleon are already relatively 
well known from deep inelastic scattering (DIS) and semi-inclusive DIS 
measurements. The gluon helicity contribution has very recently been 
found to also be nonzero 
\cite{Adamczyk:2014ozi,Adare:2015ozj,deFlorian:2014yva}, but sea quark 
helicities are still poorly understood. One of the main reasons is that 
DIS predominantly probes valence objects at the currently measured 
scales and $x$ ranges. Secondly, the uncertainties of fragmentation 
functions in semi-inclusive measurements, needed to disentangle 
different flavors, dominate the existing sea quark helicity extractions. 
An elegant alternative to access sea quark helicities is via the weak 
interaction. Such processes are possible at the high scales proposed at 
a polarized electron-ion collider \cite{Accardi:2012qut} or currently in 
polarized $p$$+$$p$ collisions at the Relativistic Heavy Ion 
Collider (RHIC) \cite{Bourrely:1993dd}. In $p$$+$$p$ collisions, real 
$W$'s can be produced in the annihilation of predominantly up and 
anti-down quark pairs for $W^+$ production and down and anti-up quark 
pairs for $W^-$ production (if one neglects the small off-diagonal 
Cabbibo-Kobayashi-Masukawa matrix elements). Furthermore, the helicity 
of participating quarks and anti-quarks is fixed to be left-handed and 
right-handed, respectively, due to the parity violating nature of the 
weak interaction. If one of the two proton beams is longitudinally 
polarized, the helicity of the proton beam therefore selects quarks that 
are polarized parallel or anti-parallel with it and vice versa for 
anti-quarks. Building the difference of the $W$ production cross 
sections for positive and negative helicities normalized by their sum, 
one arrives at the single longitudinal spin asymmetry: \begin{equation} 
A_L^{pp\rightarrow W^+} \approx \frac{ \Delta\overline{d}(x_1,Q)u(x_2,Q) 
- \Delta u(x_1,Q)\overline{d}(x_2,Q)}{ \overline{d}(x_1,Q)u(x_2,Q) + 
u(x_1,Q)\overline{d}(x_2,Q)}\quad, \end{equation} in terms of the 
unpolarized parton distribution functions (PDFs) for up and anti-down 
quarks, $u(x,Q)$ and $\overline{d}(x,Q)$, and their respective helicity 
PDFs $\Delta u(x,Q)$ and $\Delta \overline{d}(x,Q)$. The corresponding 
single spin asymmetry for $W^-$ production becomes: \begin{equation} 
A_L^{pp\rightarrow W^-} \approx \frac{ \Delta\overline{u}(x_1,Q)d(x_2,Q) 
- \Delta d(x_1,Q)\overline{u}(x_2,Q)}{ \overline{u}(x_1,Q)d(x_2,Q) + 
d(x_1,Q)\overline{u}(x_2,Q)}\quad. \end{equation} It accesses the other 
combination of light quark flavors. While the $W$ production cross 
section is relatively low compared to strong processes, the scale is set 
by the mass of the produced $W$'s. Furthermore, no uncertainties due to 
fragmentation functions enter the interpretation of these single spin 
asymmetry measurements.

In the PHENIX experiment \cite{Adcox:2003zm}, $W$'s are not 
reconstructed kinematically themselves, but their leptonic decays 
($W\rightarrow l \overline{\nu_l}$) are measured inclusively by 
detecting the charged decay lepton $l$ only. At central rapidities, $W$ 
decay electrons are reconstructed, while at forward rapidities 
($1.1<|\eta|<2.6$ and $1.1<|\eta|<2.5$ for the north and south muon 
arms, respectively) decay muons are being studied. Recent results by 
STAR \cite{Adamczyk:2014xyw} and PHENIX 
\cite{Adare:2010xa,Adare:2015gsd} for the electron channels exist. In 
this paper, the first asymmetry measurement using muons and at 
forward/backward rapidities is reported.

\begin{figure*}
\begin{minipage}{0.78\linewidth}
\includegraphics[width=0.99\textwidth]{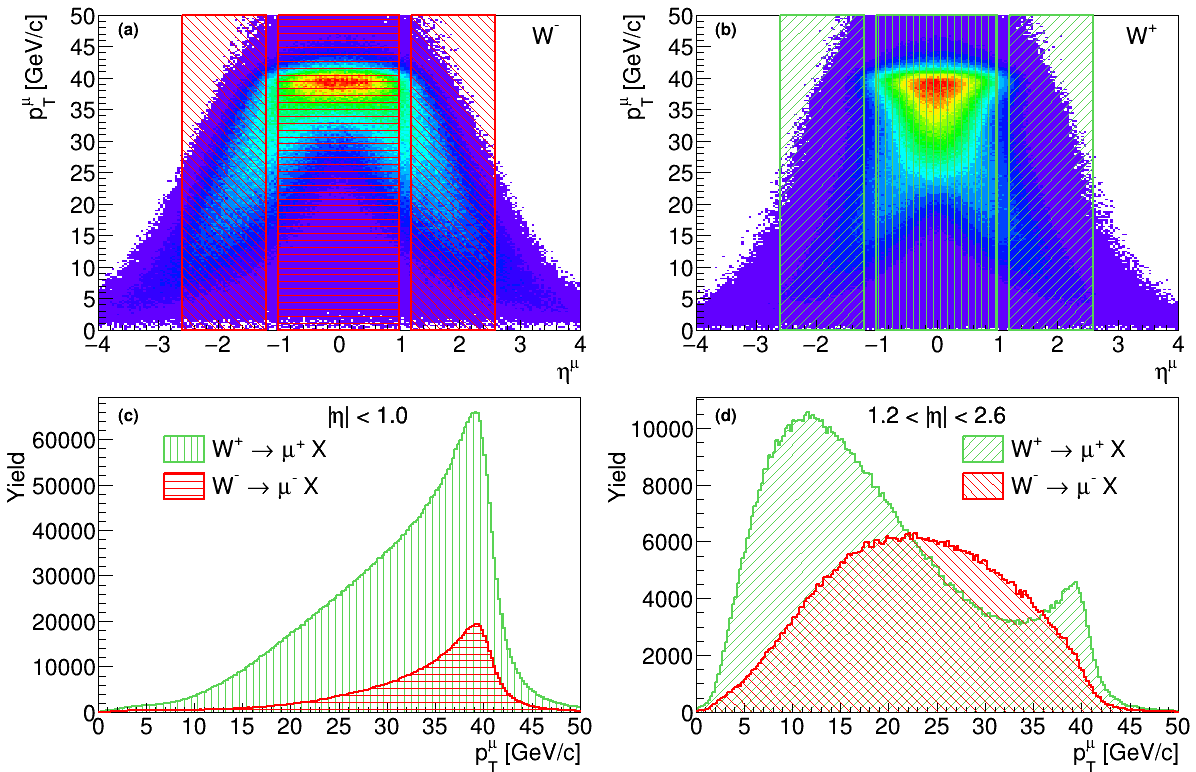}
\end{minipage}
\begin{minipage}{0.20\linewidth}
\caption{\label{fig:kinematics}
(a,b) Two-dimensional rapidity-transverse-momentum distributions for 
(a) $W^-{\rightarrow}\mu^-$ decays and (b) $W^+{\rightarrow}\mu^+$ decays. 
(c,d) Muon transverse momentum projected yields for 
(c) central rapidities ($|\eta|<1$) and (d) forward rapidities 
($1.2<|\eta|<2.6$).  The positive decay muons are displayed in the 
(c) vertical and in (d) $+45^{\circ}$ from vertical [green] hatched regions, 
while the negative decay muons are displayed in (c) horizontal 
and in (d) $-45^{\circ}$ from vertical hatched [red] regions.
}
\end{minipage}
\end{figure*}

In this analysis, we rely solely on the reconstruction of forward-going 
muons impinging the muon-spectrometer as the nonhermetic coverage of 
the PHENIX detector precludes a missing energy analysis to measure the 
neutrino. Although approximately half of the energy of the $W$ is 
carried by the muon, only a small Jacobian peak is expected in the 
forward region, in contrast to earlier measurements at midrapidity. The 
reason is the additional longitudinal momentum which takes up a 
substantial part of the $W$ decay muon's energy, as well as any nonzero 
initial $W$ momentum. The very different kinematic regimes for central 
and forward $W$ decay muons and respective yields are illustrated in 
Fig.~\ref{fig:kinematics} based on {\sc pythia}-6 simulations 
\cite{Sjostrand:2006za}. Furthermore, any remnant Jacobian peak is 
completely washed out by the limited momentum resolution of the 
muon-spectrometer at large momenta. Consequently, a data-driven approach 
has been employed to identify the contributions by the various 
backgrounds in the data sample to extract $W$ production cross 
sections and the corresponding single spin asymmetries. It should be 
noted that at forward rapidities, higher/lower $x$ of around 0.3/0.1 
from the forward/backward going proton can be probed in comparison to 
more central rapidities where both $x$ are around 0.2. For $W^-$ decays, 
the forward region also cleanly separates the down and anti-up quark 
contributions by the forward/backward going protons while for $W^+$ 
decays a mixture of up and anti-down quarks always contributes although 
at rather different $x$.

This paper is organized as follows: In section II, the different data 
taking periods and the corresponding data sets are discussed including a 
brief description of the relevant detector systems and Monte Carlo sets 
used. Section III describes the initial event selection 
criteria used to screen the raw data for events with a high likelihood 
of containing $W$ decay muons. In section IV, the extraction of the 
actual signal, the $W$ production cross section and the asymmetries are 
discussed before discussing the systematic studies in Section V. The 
corresponding results are presented in section VI before a summary of 
the measurements in the last section.

\section{Data sets}

\begin{figure*}[thb]
\begin{minipage}{0.8\linewidth}
\includegraphics[width=0.99\linewidth,trim={0 10 0 415},clip]{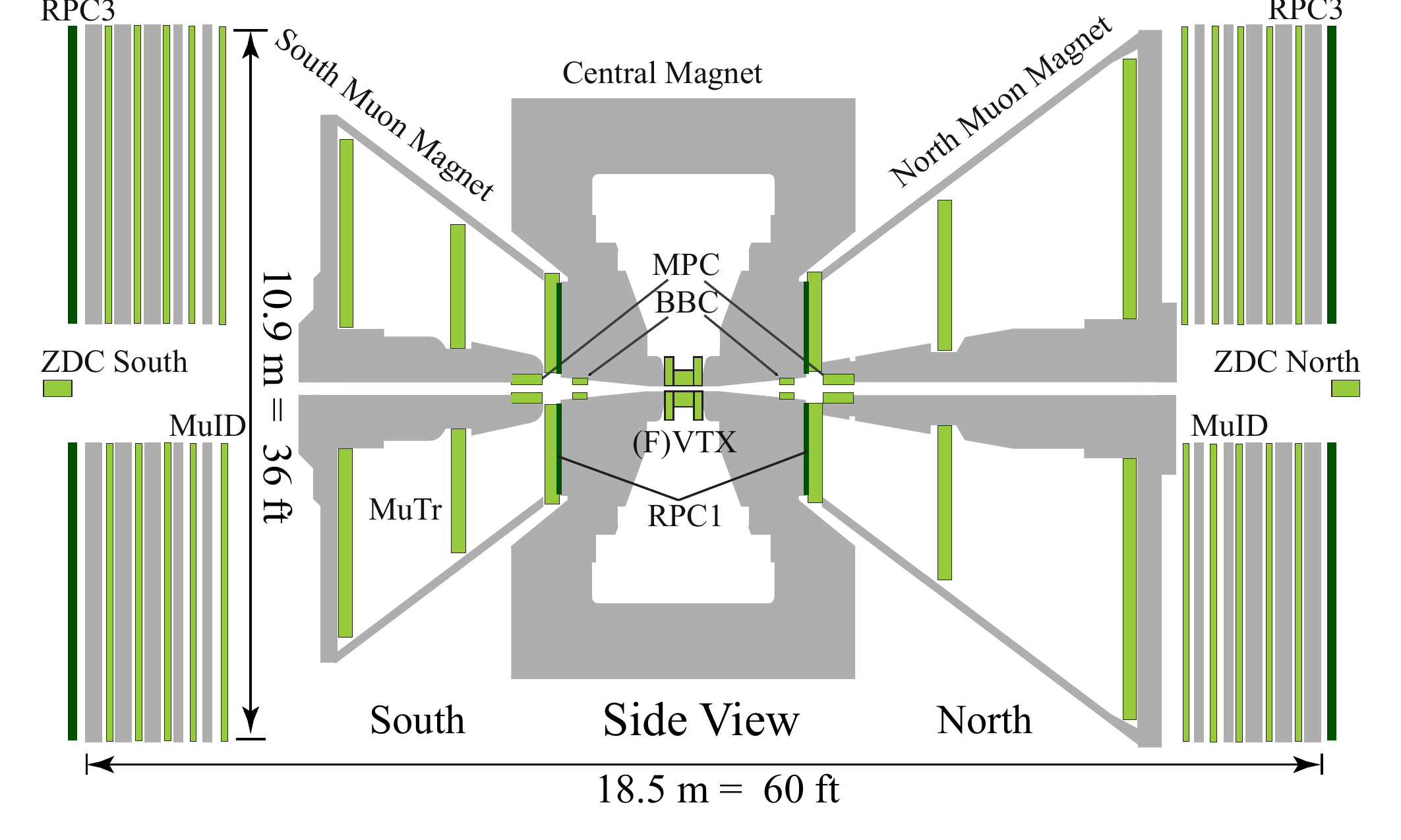}
\end{minipage}
\begin{minipage}{0.18\linewidth}
\caption{\label{fig:phenix}
A side view of the PHENIX detector, concentrating on the muon arm 
instrumentation.  Of primary importance to this analysis are the 
BBC, FVTX, RPC, MuTr, and MuID.  Please see text for descriptions
of these subsystems and how they were used.
}
\end{minipage}
\end{figure*}

The data sets used in this analysis were recorded at RHIC (Brookhaven 
National Laboratory) during the 2012 and 2013 polarized proton running 
periods at a center-of-mass energy $\sqrt{s}$\,=\, 510\,GeV. A 
luminosity of approximately 53 and 285 pb\,$^{-1}$ sampled within a wide 
vertex region of about 40 cm width was used for this analysis for the 
two running periods, accumulated with the PHENIX detector (see 
Fig.~\ref{fig:phenix}). The average beam polarizations were 56\% and 
58\% for the two beams in the 2012 running period and 54\% and 55\% in 
the 2013 running period. The polarization uncertainty was obtained by 
the RHIC polarimetry group and amounts to a relative 3\% per beam. These 
uncertainties translate into a global normalization uncertainty of the 
extracted asymmetries.

The $W$$\rightarrow$$\mu$ candidate events were detected via tracks in the 
forward muon arm system~\cite{Akikawa:2003zs}, which comprises 
the muon tracker (MuTr) and muon identifier (MuID) subsystems.
A newly installed set of resistive-plate 
chamber (RPC) detectors \cite{Meredith:2009zz}
in the muon arms, were also used to associate tracks 
with particular beam crossings and to help with triggering.  For tracks 
traversing the full array of detectors, the full azimuthal acceptance is 
covered over the region 1.1$<$$|$$\eta$$|$$<$2.5 and 
1.1$<$$|$$\eta$$|$$<$2.6 in the two arms, respectively.  However, the 
range of vertex longitudinal positions is reduced at the boundaries of 
the $\eta$ coverage.  The collision vertex was determined by the PHENIX 
beam-beam counters (BBC), which are two sets of 64 \v{C}erenkov counters 
with a pseudorapidity range of $3.0<|\eta|<3.9$.

Prior to these data sets, no momentum selectivity was available for 
triggering forward muons in PHENIX; only an enhancement of real muons 
(rejection of fake tracks) was available based on the activity in the 
downstream MuID planes. To enhance the $W$ data sample within the 
limited bandwidth available, the forward PHENIX detectors were upgraded 
to allow triggering on all $W$$\rightarrow$$\mu$ candidates. New readout 
electronics for the MuTr (MuTrig)~\cite{Adachi:2013qha} were 
added, and the RPCs were installed upstream and downstream of the Muon 
arms. In both subsystems, the azimuthal segmentation allowed for the 
selection of events with muon candidates traversing the whole muon 
system and with nearly straight lines in real time. Depending on the 
polar angular coverage of the RPCs, three main types of triggers were 
created.  At low pseudorapidities ($|\eta|<1.4$), only the upstream RPCs 
were in coincidence with the MuTrig and collision counters, while at 
high pseudorapidities ($|\eta|>2.0$), only the downstream RPCs were 
available for coincidence with the MuTrig and collision counters. In the 
intermediate region, a coincidence of both RPCs and the MuTrig was 
required.  This new trigger selected track candidates that satisfied a 
minimum momentum threshold of approximately 10\,GeV/$c$.

Several other trigger combinations less sensitive to momentum were 
considered at reduced data taking rates for this analysis, in addition 
to the main triggers described above. In 2012 only the downstream RPCs 
were part of the trigger while the upstream RPCs were only used in the 
offline analysis. For triggers without downstream RPC information a 
coincidence with a hit in the furthest plane of the MuID is required, 
enforcing the track candidate to penetrate at least 12.8 $\lambda_{I}$ 
(1.1$<$$\eta$$<$2.6) and 12.0 $\lambda_{I}$ ($-$2.4$<$$\eta$$<$$-$1.1) 
nuclear interaction lengths. For a trigger cross-check and efficiency 
evaluation, independent data samples were collected which only relied on 
the muon identifier or entirely different PHENIX detector components.

\begin{figure}[thb]
\includegraphics[width=1.0\linewidth]{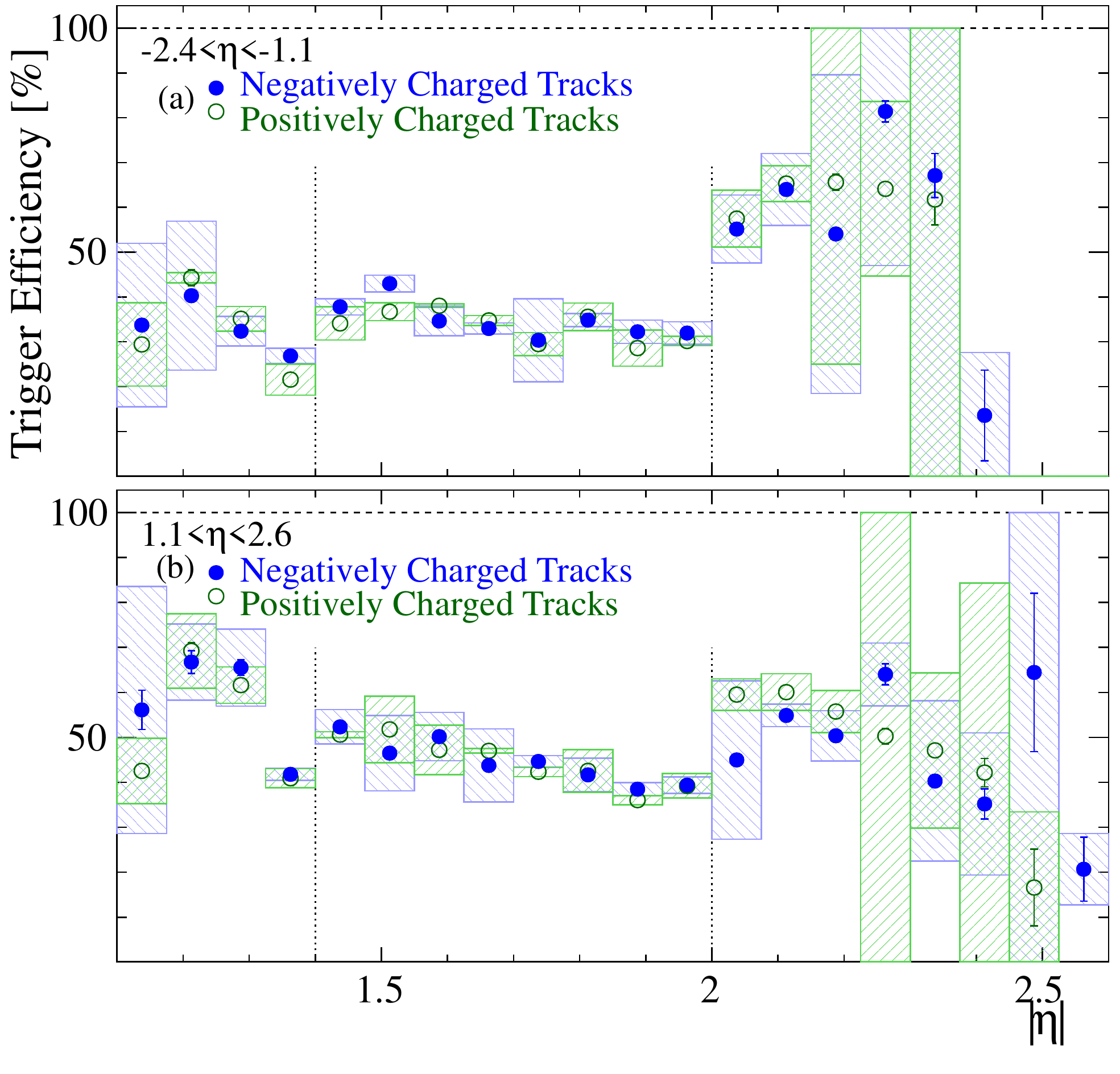}
\caption{\label{fig:trigeffi}
Trigger efficiencies for the 2013 running period as a function of $\eta$ 
for negative (solid circle [blue]) and positive (open circle [green]) $W$ 
decay muon candidates in (a) south and (b) north PHENIX muon detector arms 
for pseudorapidity ranges (a) $-2.4<{\eta}<-1.1$ and (b) $1.1<{\eta}<2.6$.  
The hatched boxes correspond to the systematic uncertainties in the trigger 
efficiency calculations.  The vertical, dotted lines represent the 
approximate boundaries of the three main RPC-based triggers.
}
\end{figure}

As the data collision rate far exceeded the capacity of the PHENIX data 
acquisition system to record data, only a small fraction (1 in every 30 
to 130 events, depending on luminosity) of this data was written to tape 
for further analysis, while for the new momentum-sensitive triggers, 
essentially all events were recorded. The total trigger efficiencies for 
$W$-decay muon candidate events varied as a function of rapidity due to 
the combination of different trigger components according to their 
individual ranges of coverage. For example, at very low (high) 
rapidities, only upstream (downstream) RPCs were available, which 
reduced their rejection rates but increased the trigger efficiencies in 
these regions. These trigger efficiencies are summarized in 
Fig.~\ref{fig:trigeffi} for positive and negative muons at forward and 
backward rapidities, showing the discussed rapidity dependence for the 
2013 data taking period. In the 2012 running period, the upstream RPCs 
were still being commissioned, and simpler triggers using mostly the 
MuTrig information, the MuID, and the downstream RPCs were used. As 
such, the rapidity range was more limited, but the efficiencies were 
nearly constant over that range (approximately 50\% to 60\% for the two 
arms and charges).

Moreover, in the 2012 and 2013 running periods, a new forward vertex 
detector (FVTX) was available \cite{Aidala:2013vna}, consisting of 4 
planes of Silicon strips finely segmented in radius and coarsely 
segmented in azimuth. For the subset of muon candidate tracks passing 
several of these detector planes (about 10\%--30\% of tracks), this 
additional information was used to improve tracking quality and to 
further reduce jet-like events.

During collider downtime and periods prior to and after the end of the 
2011-2013 physics runs, cosmic-ray data was collected.  The rate of 
high-energy cosmic-ray muons in the PHENIX detector -- as a potential 
background to the $W$ signal -- was found to be negligible (below 1\% of 
the expected W decay muons) when applying the same selection criteria as 
for $W$ decay signal candidates.  This large sample of cosmic-ray events 
provides crucial information on the reconstruction performance of the 
muon arms for high-momentum tracks.  Muons traversing both spectrometer 
arms are reconstructed as a pair of back-to-back muon tracks that have 
nearly the same momenta but opposite charge sign. Incoming tracks 
enter the spectrometer arm from outside the detector volume, pass 
through the detector, and exit through the opposite spectrometer. They 
are also required to pass the nominal vertex region, which, together 
with the two-arm requirements, limits their $\eta$ acceptance in 
comparison to the W analysis.

\begin{figure}[thb]
    \includegraphics[width=1.0\linewidth]{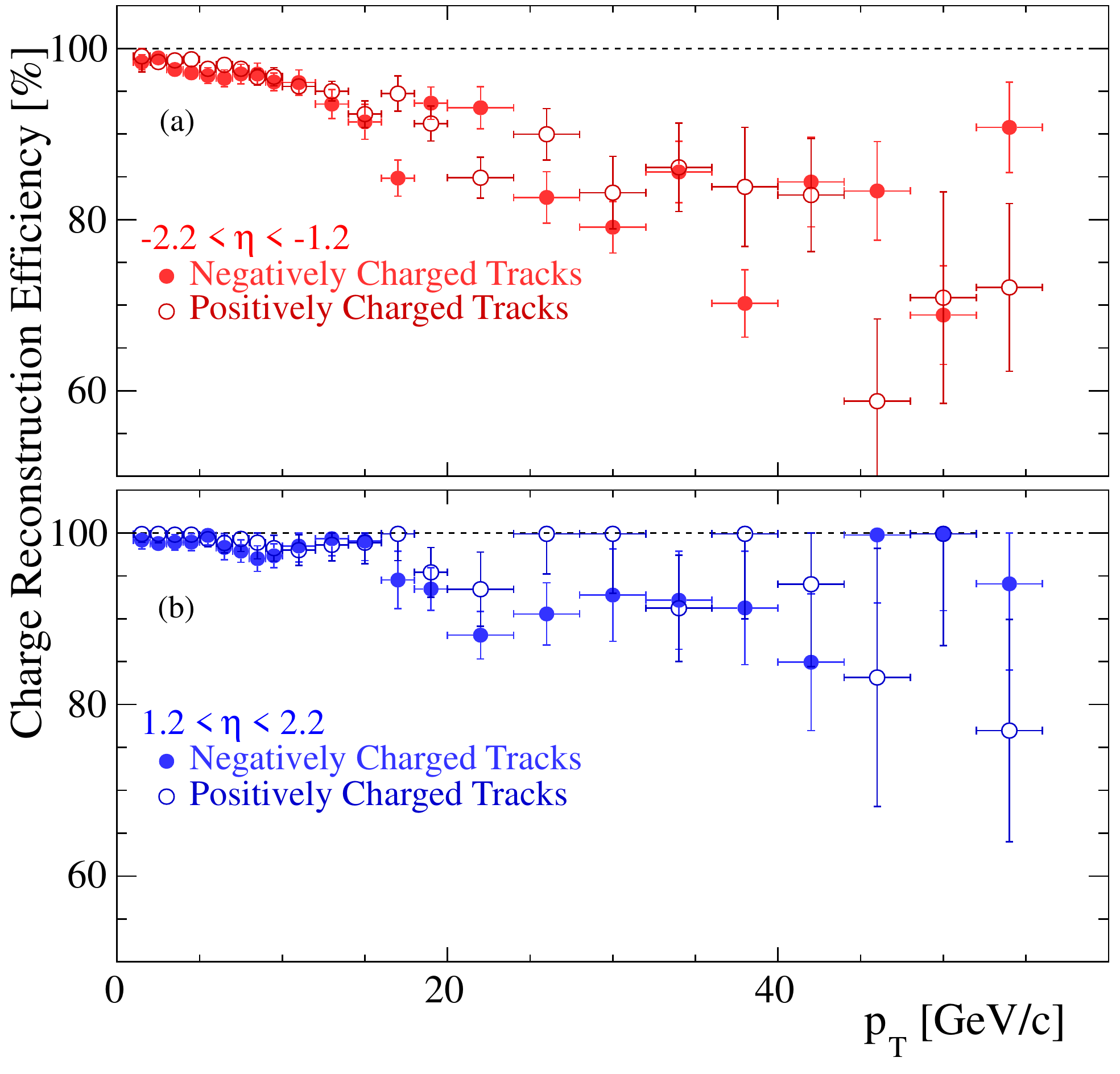}
    \caption{\label{fig:CosmicChSign} 
Charge-sign reconstruction-efficiency for the (a) south and (b) north muon 
arms.  The (a) $-2.2<{\eta}<-1.2$ [red] and (b) $1.2<{\eta}<2.2$ [blue] 
solid and open circles are for negatively and positively charged 
tracks, respectively.
}
\end{figure}

Using this data, the charge sign reconstruction efficiencies were 
investigated. Owing to the limitations in the spectrometer segmentation, 
measuring the bend plane becomes ambiguous for the highest momentum 
tracks, which are almost straight.  The rate of each incoming charge 
sign is compared to the rate of oppositely-charged outgoing muons.  The 
difference is an inefficiency in the charge sign reconstruction.  The 
results of this test are shown in Fig.~\ref{fig:CosmicChSign}.  As 
expected, the charge sign reconstruction is $\sim$100\% for low momentum 
tracks and is $\sim$90\% ($\sim$80\%) for the spectrometer located at 1.1 
$<$ $\eta$ $<$ 2.6 (-2.5 $<$ $\eta$ $<$ -1.1) at high momenta. These 
results are found to be well reproduced in simulations.

\begin{figure}[htb]
    \includegraphics[width=1.0\linewidth]{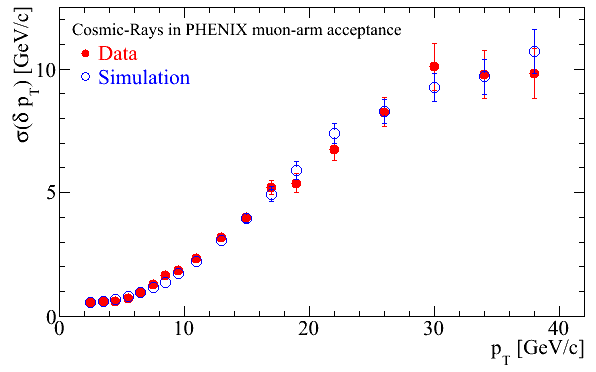}
    \caption{\label{fig:CosmicSmearing} 
Momentum reconstruction resolution from measured (solid [red] circles) 
and simulated (open [blue] circles) cosmic-ray muons.
}
\end{figure}

Direct comparison of the reconstructed momentum from the incoming and 
outgoing part of each cosmic muon track indicates the accuracy of the 
momentum reconstruction. Fig.~\ref{fig:CosmicSmearing} shows the 
relative (between arms) resolution of the transverse momentum 
reconstruction for cosmic data and simulation.  Although the accuracy is 
low for high-momentum tracks ($\sigma$$\sim$25\%), this is well 
reproduced in the simulations, indicating that the data will be 
accurately imitated by the simulations. The variation of this momentum 
reconstruction accuracy in simulations will be considered as 
uncertainties due to detector smearing. An additional rate-dependent 
degradation in the momentum smearing was taken into account in the 
corresponding simulations.

In addition to the collected physics data sample, several sets of 
Monte-Carlo (MC) simulated data were produced and analyzed. First, a 
large sample of {\sc pythia}+{\sc geant3} \cite{Brun:1994aa} simulations 
was used to estimate the reconstruction efficiency for high-momentum 
muons representing the $W$ signal. In addition, {\sc rhicbos} 
\cite{Nadolsky:2003ga} and {\sc che} \cite{deFlorian:2010aa} were used 
as generators for signal events. Muonic decays of $Z$ boson production 
were included in the signal simulations as they are indistinguishable in 
this analysis.  To ensure that the MC represents the collected data, 
residuals between hits and reconstructed tracks and other kinematic 
distributions were compared to pure muon samples collected from 
cosmic-ray and positively-identified muons (from $J/\psi$ decays).  The 
MC distributions were found to reproduce those in data accurately.

Similar large scale {\sc pythia}+{\sc geant} simulations were also 
performed for various background contributions. Heavy flavor decays into 
muons as well as muonic decays of charmonium and bottomonium resonances 
dilute the $W$ decay muon signal as they are indistinguishable from real 
$W$ boson decay muons. Due to the sizable momentum smearing at high 
reconstructed transverse momenta, these decays do contribute 
substantially even though their actual transverse momenta drop rapidly. 
To ensure that these simulated single muon background contributions 
correctly describe the real muon background, their relative 
contributions were evaluated using fits to opposite sign dimuon 
invariant mass data. The weighted simulated real muon background 
contributions were then fixed in the $W$ signal fits of the single muon 
candidates, while the individual weights' correlated uncertainties were 
assigned as systematic uncertainties.

Another suite of MC probed the contribution of background particles that 
may masquerade as $W$ decay muons. This comprises contributions from 
$\pi^{\pm}$ and $K^{\pm}$ decays. Single $\pi^{\pm}$ and $K^{\pm}$ + 
{\sc geant} simulations are used to estimate the fake background from 
in-flight decays within the muon-spectrometer. Their generated 
contributions are weighted based on next-to-leading order (NLO) 
calculations \cite{werner} in 
the same rapidity range. They are consistent with the available 
experimental data, as well as {\sc pythia} TuneA simulations that 
generally reproduce hadronic cross sections at RHIC energies. Due to 
their large initial cross section, especially at low transverse momenta, 
a sizable contribution of hadrons survive all absorbing material 
upstream of the muon tracker and their subsequent decays may appear as 
near-straight tracks and get mis-reconstructed as high-momentum tracks. 
A substantial part, more than 98\%, of such fake muons can be rejected 
due to the large amount of multiple scattering in the absorbing matter 
and thus poorer correlations between different detector systems. 
However, not all such candidates can be removed, leaving these as the 
most important background in the $W$ measurements. The majority of 
effort in this analysis concentrates on reducing this hadronic 
background, ensuring its reliability from the simulations to the data 
and fitting its contributions in the signal enhanced data sample.

\section{Event and Track selection criteria}

The triggered data sample was further analyzed to reduce contamination 
of nonmuon particles in the sample. A candidate muon was formed from 
two pieces of information: a formed track in the muon-spectrometer and a 
short ``road'' through the whole of the MuID.  Muon-like track quality 
was determined through residual distributions of track and road 
variables that combine to form a powerful method to distinguish real and 
fake muons.  These residuals can be classified into three broad 
categories: identification, track/road matching, and physics.

For each high-momentum track candidate ($16 $ GeV/c $< p_T < 60$ GeV/c) in 
the spectrometer ($1.1 < | \eta| < 2.6$) , the difference between the 
measured hit positions of the track and the subsequent fit are used to 
form a $\chi^{2}$ per degree of freedom residual. This track-$\chi^{2}$ 
residual, coupled with the requirement that the track passes through the 
whole MuID, establishes tracks as muon candidates.

The second category, track/road matching, is an ensemble of variables 
that are sensitive to differences between real muons and fake-muon 
backgrounds formed from the decay of light hadrons (particularly 
$K^{\pm}$) within the volume of the muon-spectrometer.  Such decays 
produce a kink in their track, changing the trajectory measured in the 
spectrometer relative to that in the MuID. Therefore, the angular and 
spatial differences between the track and the road of the candidate are 
wider for hadron-decay muons than for muons originating at the collision 
vertex. Also, the distribution of the projections to the collision 
vertex is broader due to the multiple scattering in the absorbing 
materials. These decay hadrons have two properties.  First, these are 
typically low-momentum hadrons that have punched through the central arm 
magnet return yoke (4.9 $\lambda_{I}$ steel) and absorber (2.3 
$\lambda_{I}$ steel) nuclear interaction lengths.  Second, the decay 
kinematics for some of these hadrons result in a mismeasurement of the 
track momentum, promoting the originally low momentum particle to higher 
momenta.  Although there is only a tiny probability of this confluence, 
the large number of light hadrons produced in soft $p$$+$$p$ interactions 
makes this the dominant source of fake-muon background in this analysis.

The final track residual category utilizes the newly installed RPC 
detectors to associate tracks with particular beam-crossings. Typically 
107 to 111 of 120 bunches were filled during these running periods with 
bunch crossings every 106 ns. The RHIC accelerator provides alternating 
orientation of the proton polarization in two groups of four 
combinations. This alternating approach minimizes systematic effects of 
individual bunch crossings' varying beam luminosity and polarization. As 
a consequence, specific tracks have to be matched in time to particular 
beam-crossings.  The RPCs provide a space-time stamp for each candidate 
track, whereby the spatial information is used to assign an RPC cluster 
to the track, and the corresponding time is used to identify the correct 
beam-crossing.  A tight requirement is imposed on the distance of 
closest approach (between the RPC cluster and the projected track 
trajectory onto the RPC plane), along with a stringent time window to 
reject tracks from prior/subsequent crossings.

Finally, matching of the fully-formed muon candidate to the collision 
vertex position (estimated using the BBC) rejects background tracks that 
do not originate at the point of collision.  For data taking in 2013, 
the RPCs and their matching information were already implemented as part 
of the main trigger while for 2012 this matching needed to be performed 
offline for the upstream RPCs.

For the FVTX detector, similar matching variables were used if several 
FVTX planes were hit and formed a FVTX track candidate. Additionally, 
for each track in the muon arms, the number of FVTX track candidates in 
the vicinity is counted. This provides additional information to 
suppress both heavy flavor and fake muon backgrounds because their muon 
candidate tracks are more likely found within a jet of particles.

\begin{figure}[tbh]
    \includegraphics[width=1.0\linewidth]{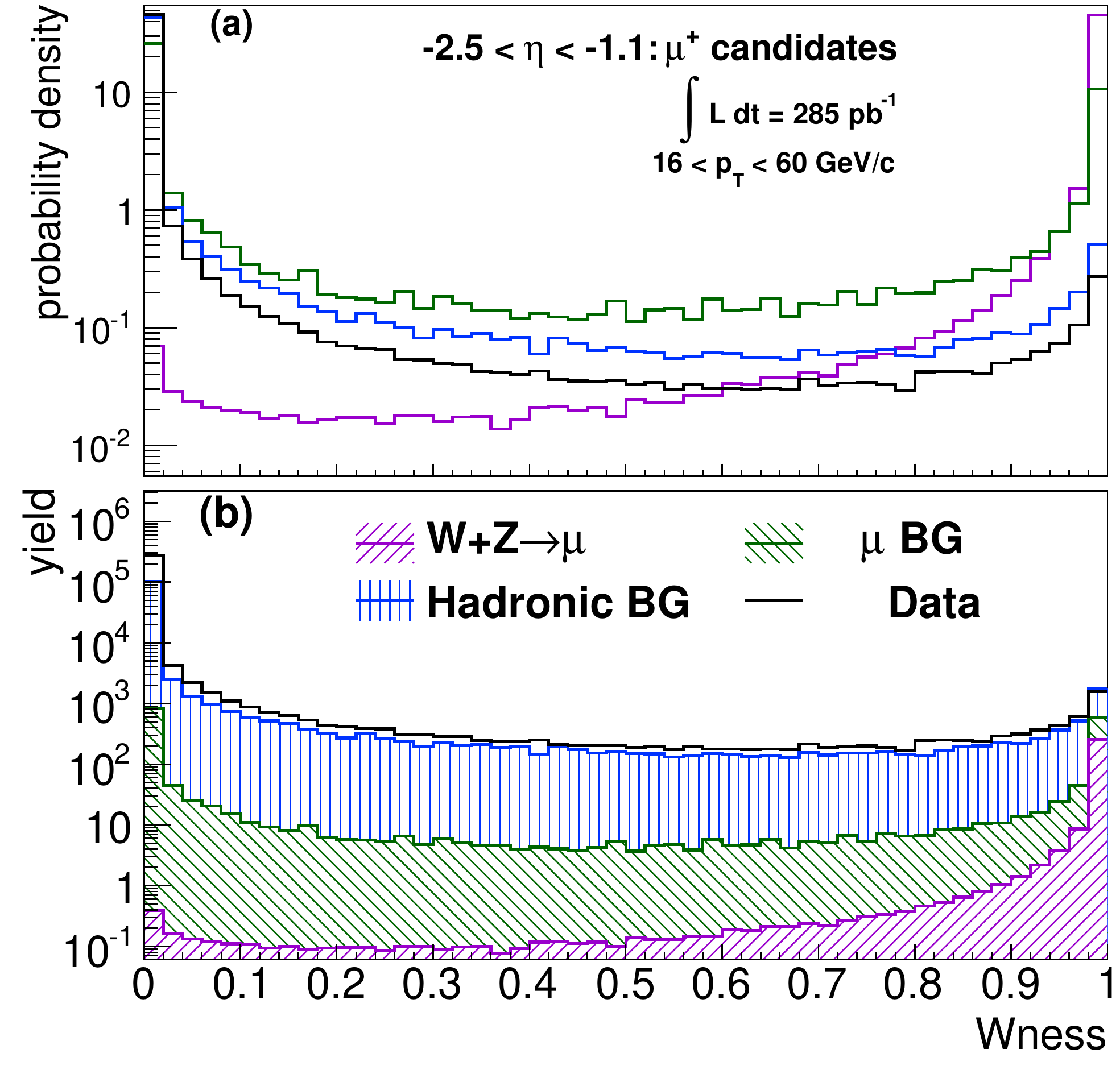}
    \caption{\label{fig:Wness}
(a) Probability density distribution for $W_{\rm ness}$ and (b) 
absolute yields of data and stacked simulated signal and background 
contributions as a function of $W_{\rm ness}$.  (a) The sequence of 
curves from top to bottom for $W_{\rm ness}<0.6$ is muon background 
[dark green], hadronic background [blue], data [black], and 
simulated signal [purple]. (b) The sequence of curve and shadings 
from top to bottom is data [black], hadronic background [blue], muon 
background [green], and simulated signal [purple].
}
\end{figure}

For each event, the value of each track residual is given a probability 
based on reference distributions from simulated $W$ decays, 
$\lambda_{Sig}$, and collected data, $\lambda_{Bg}$. The latter is 
effectively a background distribution due to the low percentage of 
signal present ($<$0.1\%). The use of the full data distribution for the 
background allows for the correct mixing of hadronic and muon 
backgrounds. A combined probability distribution, \begin{equation} 
W_{\rm ness} = \frac{\lambda_{Sig}}{\lambda_{Sig} + \lambda_{Bg}} \quad, 
\end{equation} is formed from all variables available, including track 
and road matching position and angular residuals, transverse distance to 
the vertex point, residuals to the RPC clusters, FVTX matching 
residuals, and FVTX track candidate multiplicity. It is displayed in 
Fig.~\ref{fig:Wness}, where probabilities close to unity represent 
$W$-muon like tracks, while near zero probabilities represent hadronic 
background dominated events.  Tracks with high $W_{\rm ness}$ ($>$0.92) are 
used for further analysis. This value was chosen as a compromise between 
signal purity (around 10\% to 17\%) and efficiency (above 95\%)  
to optimize the uncertainties of the background corrected asymmetries. 
The $W_{\rm ness}$ data distribution is reasonably well described by a 
combination of the individual MCs for the signal, real muon, and 
hadronic backgrounds.

As not all detector components (upstream, downstream RPCs, and FVTX 
detectors) cover the whole rapidity range, between five to nine 
kinematic and residual variables entered the combined probability. For 
correlated variables, the initial probability density functions were 
evaluated together. The different variables are summarized in Table 
\ref{tab:var}.

\begin{table*}
\caption{\label{tab:var} 
Kinematic variables used in the $W_{\rm ness}$ evaluation. Variables in 
square brackets were not available for all events due to the different acceptances 
of the respective subsystems.
}
\begin{ruledtabular} \begin{tabular}{ccrlcc}
&& DG0 &  track-road difference at first MuID plane's $z$ position && \\ 
&& DDG0 &  track-road angular difference at first MuID plane's $z$ position && \\  
&& DCA$_r$ &  radial distance of extrapolated track at vertex $z$ position && \\  
&& FVTX $N_{\rm clus}$ &   FVTX track multiplicity in cone around extrapolated track candidate && \\ 
&& RPC1$_{\rm DCA}$ &  [RPC1 hit cluster-track difference at RPC1 $z$ position] &&  \\  
&& RPC3$_{\rm DCA}$ &  [RPC3 hit cluster-road difference at RPC3 $z$ position] && \\  
&& FVTX $\Delta \phi, \Delta r$ and $\Delta \theta $ & [FVTX track and MuTr residuals] && \\ 
\end{tabular} \end{ruledtabular}
\end{table*}

\section{Signal Extraction and Backgrounds}

\begin{figure*}[thb]
\includegraphics[width=0.94\linewidth]{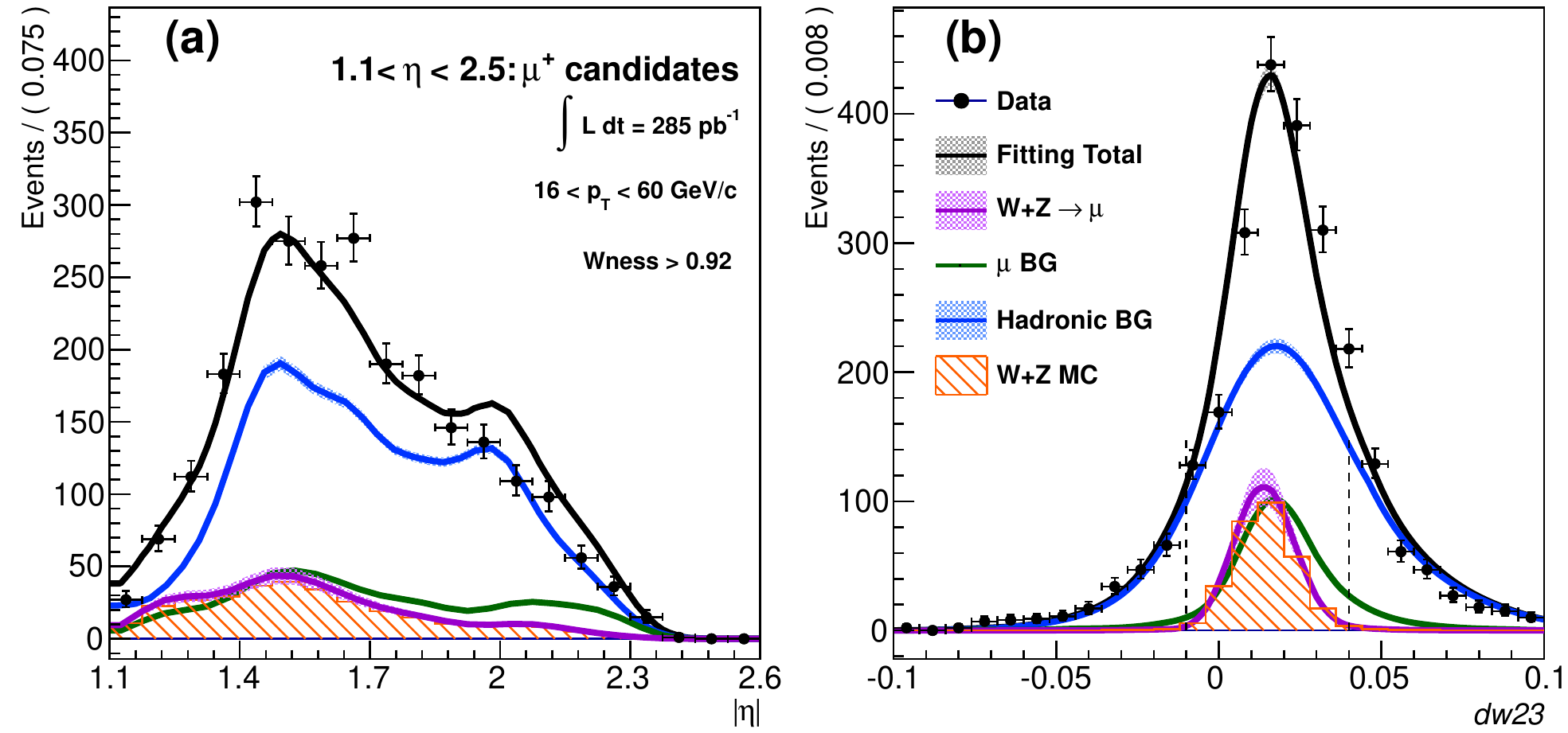}
\caption{\label{fig:SigBackFit} 
Signal and background discrimination of positive rapidity $\mu^{+}$ 
candidates.  (a) pseudorapidity and (b) azimuthal bend-plane-variable, 
$dw_{23}$, dependence from top to bottom for data points, curves, and 
hatching for (a) $|\eta|=$1.6--2.3 and (b) $dw_{23}=$0.024--0.04 are data 
(solid [black] circles), combined fitting total [black], hadronic 
background [blue], real-muon background [dark green], signal [purple], and 
MC-based expected-signal yields (hatched [red]). The relative yields of 
signal and backgrounds are determined from a simultaneous fit to both 
discriminant distributions. The additional azimuthal bend-plane variable 
range selection for the asymmetry evaluation is shown as vertical dashed lines.
}
\end{figure*}

After selecting candidate tracks as muon candidates, most of the 
remaining tracks are still not muons from $W$ decays. An unbinned 
maximum likelihood fit approach is used to determine the final number of 
$W$s and remaining backgrounds. Figure~\ref{fig:SigBackFit} illustrates 
the two discriminant variables used to normalize the relative 
contribution of the signal and backgrounds.  
Figure~\ref{fig:SigBackFit}(a) shows the pseudorapidity of the track for 
data (solid circles) along with the expected distributions from signal 
muons (purple solid line), background muons (green solid line), and the 
residual mis-identified hadronic background (blue solid line). 
Figure~\ref{fig:SigBackFit}(b) shows a variable determined from the 
azimuthal bend-plane between the second and third muon spectrometer 
stations, $dw_{23} = \Delta \phi_{23} \times \sin\theta \times p_{T}$, 
where $\Delta \phi_{23}$ is the difference of the azimuthal angle 
between the second and third station hits and $\theta$ is the polar 
angle of the track relative to the beam direction. The pseudorapidity 
and $dw_{23}$ variables are found to be almost orthogonal in 
sensitivity.

The underlying discriminant shapes for the signal $W$-muons and real 
muon backgrounds are determined from the MC simulations. The 
normalization (for the muon backgrounds) is determined from the yield of 
$c$, $b$, and quarkonia decays in {\sc pythia} as evaluated via fits to 
the dimuon data. For the hadronic background shape, the azimuthal 
bend-plane distributions were extracted from hadron simulations directly 
in the target $W_{\rm ness}$ region ($W_{\rm ness}>0.92$) as no unbiased hadronic 
background only data sample was available. The pseudorapidity variable 
is extracted from data in the target $W_{\rm ness}$ region from the side bands 
of the azimuthal bend-plane variable where neither signal nor background 
muons contribute. It corresponds to $dw_{23} < -0.05 (-0.01)$ or 
$dw_{23} > 0.01(0.05)$ for negative (positive) tracks, respectively. 
Final signal-to-background ratios vary from 10\% to 17\%, depending on 
charge sign and spectrometer arm before restricting the azimuthal 
bend-plane variable for the asymmetry analysis. To obtain the 
corresponding cross sections, the extracted signal yields get corrected 
for charge mis-identification ($<$2\%), reconstruction and acceptance 
efficiencies (approximately 0.6\% for $W^+$ and 2\% for $W^-$), trigger 
efficiencies and, Z boson admixture (18\% to 22\%). The yields are then 
normalized by the accumulated luminosity to arrive at the 
$W$$\rightarrow \mu$ cross sections.

To extract the single spin asymmetries, the high $W_{\rm ness}$ 
($W_{\rm ness}>0.92$) data sample was taken with the additional selection of 
the azimuthal bend-plane variable with $W$ support only ($\mp 0.01$ to 
$\pm 0.04$, for positive and negative charges, respectively) and 
rapidities $\eta < 2$. In this region, the signal-to-background ratios 
increase to between 15\% to 28\%. The yields were separated according to 
the helicity combinations normalized by the corresponding beam 
polarizations. For each arm and charge, a single spin asymmetry for each 
beam and a combined double spin asymmetry can be extracted. The 
differences in relative luminosity were accounted for by using scalers 
from the PHENIX collision counters as relative weights. The 
uncertainties on these correction factors are insignificant relative to 
the other uncertainties. No background process should possess a 
parity-violating asymmetry. This was experimentally verified by either 
selecting muon candidates at lower transverse momenta or lower $W_{\rm ness}$. 
Consequently, the actual $W+Z$ single spin asymmetries can be extracted 
from the raw asymmetries by correcting the dilution from the background 
using the obtained signal-to-background ratios. Because the 
signal-to-background ratios are still well below unity, the variation of 
the background correction, according to the uncertainties on the 
signal-to-background ratios, results in large systematic uncertainties 
on the asymmetries, which are comparable to the statistical 
uncertainties.

\section{Systematic studies}

To estimate the systematic uncertainties on the 
signal-to-background ratios, several factors impacting them have been 
varied. The most important factor involves the amount of real muon 
backgrounds because that contribution was fixed in the unbinned maximum 
likelihood fits. As real muon backgrounds most closely resemble the 
rapidity and azimuthal bend-plane distribution of the signal, a 
smaller/larger muon background weight gets preferentially compensated 
with a larger/smaller signal yield. To evaluate these uncertainties, the 
amount of muon backgrounds was varied according to the uncertainties 
obtained on the individual weights of the various charm and bottom 
contributions in the dimuon fits. As some of these values are 
correlated, the calculated correlation matrix was fully taken into 
account when varying these contributions. Another uncertainty originates 
from varying the trigger efficiency, which affects the real muon 
backgrounds, as well as the total reconstruction efficiencies for the 
cross section measurements.  While the effect of varying the trigger 
efficiencies according to their uncertainties is small, in the 
reconstruction efficiency correction, it again enters the 
signal-to-background fits via the size of the muon backgrounds. The 
trigger efficiencies were varied according to their statistical and 
systematic uncertainties. The latter originates from different ways of 
extrapolating the trigger efficiencies to the high $W_{\rm ness}$ region, as 
well as from the use of different reference data samples to obtain the 
trigger efficiencies.  Another uncertainty affecting the real muon 
backgrounds, as well as the total cross section, is the uncertainty on 
the accumulated luminosity. A dedicated analysis using van der Meer 
scans to obtain the total cross sections for the luminosity detectors in 
PHENIX determined the systematic uncertainty to be 10\%. The luminosity 
has been varied accordingly in fit and cross section calculation to 
obtain the corresponding uncertainty. The correctness of the signal 
extraction procedure is tested in fully MC simulated data. While 
generally found reliable, a tendency of the hadron background shape 
extraction to cause the signal to be overestimated in the fits was 
found. As a consequence, a systematic uncertainty is assigned according 
to the relative overestimation seen in these fully simulated MC fits. 
Additional uncertainties are obtained by varying the momentum smearing 
in the signal and background simulations according to the experimentally 
found uncertainties.

To obtain cross sections, the extracted yields need to be 
normalized by the accumulated luminosity and corrected for 
reconstruction efficiencies and acceptance. For the reconstruction 
efficiency and acceptance correction two methods were used. Either the 
signal from {\sc pythia} events were used to evaluate the correction 
factor or from the NLO generator {\sc rhicbos} \cite{Nadolsky:2003ga}. 
In both cases, the rapidity dependence is quite similar, and the 
differences were assigned as systematic uncertainties. Also, the 
dependence on the collision rate has been taken into account. Similarly, 
we cannot experimentally identify and remove Z boson decays to muons, so 
we used these two MC generators to remove the Z contributions. These 
contributions amount to about 18\% to 22\% for positive and negative 
muons, respectively. Again, the differences between {\sc pythia} and 
{\sc rhicbos} were assigned as systematic uncertainties. As was shown in 
Fig.~\ref{fig:CosmicChSign}, the charge reconstruction efficiencies are 
generally very high and well described by MC simulations. The 
efficiencies are found to drop towards low absolute pseudorapidities. To 
estimate its possible effect, the difference from the results with a charge 
mis-identification rate of 20\% was assigned as systematic uncertainty. 
Due to the larger yields for positive muons, this systematic uncertainty 
results in lower uncertainties on the $W^-$ and upper uncertainties on 
the $W^+$ cross sections.

All these contributions were varied either in the unbinned maximum 
likelihood fits directly or in the cross section extractions. The 
individual uncertainties were assumed to be uncorrelated, and a Gaussian 
sampling technique was applied to obtain the total uncertainties on the 
signal-to-background ratios as well as for the cross sections. For the 
asymmetry calculations the uncertainties on the signal-to-background 
ratios as well as the impact of charge mis-identification and smearing 
were again taken into account in the background-corrected asymmetries. 
The systematic uncertainties of the cross section measurements are 
summarized in Table \ref{tab:xsecsystematics}.


\begin{table*}[tbh] 
\caption{\label{tab:xsecsystematics} 
$W\rightarrow \mu^\pm$ cross section systematic table for the 2013 data 
in pb. The uncertainties of the 2012 data set are comparable. The 
individual contributions and their asymmetric lower and upper systematic 
uncertainties, denoted as lower and upper, are given for each charge and 
arm.
}
\begin{ruledtabular} \begin{tabular}{cccccccccc} 
           & \multicolumn{4}{c}{south muon arm} 
         & & \multicolumn{4}{c}{north muon arm} \\
           & \multicolumn{2}{c}{$W^-{\rightarrow}\mu^-$} 
           & \multicolumn{2}{c}{$W^+{\rightarrow}\mu^+$} 
         & & \multicolumn{2}{c}{$W^-{\rightarrow}\mu^-$} 
           & \multicolumn{2}{c}{$W^+{\rightarrow}\mu^+$} \\
Systematic & lower & upper & lower & upper 
         & & lower & upper & lower & upper \\ 
\hline
Smearing   
           &   4.12 &  2.48  & 14.67 & 11.16  
         & &   1.65 &  1.87  &  5.89 & 7.90 \\ 
$\mu$ BG   
           &  13.60 & 13.71  & 33.82 & 33.80
         & &  11.51 & 11.66  & 24.76 & 24.87 \\
MC checks  
           &   4.06 &  0.00  & 18.22 &  0.00
        &  &  11.23 &  0.00  & 12.94 &  0.00 \\
Trigger efficiencies 
             & 2.37 &  0.56  &  4.63 &  4.11
        &    & 1.81 &  1.59  & 2.93 & 2.73 \\
Luminosity scale  
             & 0.09 &  0.07  & 6.53 & 8.00 
          &  & 1.51 &  1.85  & 4.64 & 5.67 \\
Charge reconstruction efficiency 
           &   9.67 &  0.31 & 1.04 & 31.59
       &   &   0.08 &  6.00 & 18.43 & 0.28 \\
$Z$ admixture 
           &   1.50 &  0.00  &  0.04 &  0.00 
        &  &   2.32 &  0.00  &  0.41 &  0.00 \\ 
Acceptance  
        &      1.83 &  4.44  &  8.93 & 21.91 
    &      &   2.31 &  3.26  &  5.63 &  9.92 \\
\end{tabular} \end{ruledtabular}
\end{table*} 

\begingroup \squeezetable
\begin{table*}[tbh]
  \caption{\label{tble:Systs} 
Single-spin asymmetries at forward $A^{FW}_L$ and backward $A^{BW}_L$ rapidities 
for $p$$+$$p$ collisions at $\sqrt{s}=510$ GeV 
for results in 2013 and 2012, plus combined results for both years.
}
  \begin{ruledtabular} \begin{tabular}{ccccc}
Year & $A^{FW}_L(W^++Z \rightarrow \mu^+)$ & $A^{BW}_L(W^++Z \rightarrow \mu^+)$ 
     & $A^{FW}_L(W^-+Z \rightarrow \mu^-)$ & $A^{BW}_L(W^-+Z \rightarrow \mu^-)$ \\
\hline
2013 & $ -0.252 \pm 0.18 (\rm stat) ^{+0.18} _{-0.24} (\rm syst)$ 
     & $ \phantom{-}0.097 \pm 0.18 (\rm stat) ^{+0.21} _{-0.16} (\rm syst)$ 
     & $ -0.057 \pm 0.18 (\rm stat) ^{+0.31} _{-0.32} (\rm syst)$ 
     & $ \phantom{-}0.201 \pm 0.18 (\rm stat) ^{+0.38} _{-0.31} (\rm syst)$  \\ 
2012 & $ -0.321 \pm 0.22 (\rm stat) ^{+0.28} _{-0.36} (\rm syst)$ 
     & $ \phantom{-}0.107 \pm 0.22 (\rm stat) ^{+0.29} _{-0.24} (\rm syst)$ 
     & $ -0.153 \pm 0.39 (\rm stat) ^{+0.53} _{-0.44} (\rm syst)$ 
     & $ \phantom{-}0.481 \pm 0.37 (\rm stat) ^{+0.41} _{-0.65} (\rm syst)$  \\
Both & $ -0.283 \pm 0.14 (\rm stat) ^{+0.23} _{-0.29} (\rm syst)$ 
     & $ \phantom{-}0.102 \pm 0.14 (\rm stat) ^{+0.24} _{-0.20} (\rm syst)$ 
     & $ -0.087 \pm 0.16 (\rm stat) ^{+0.38} _{-0.35} (\rm syst)$ 
     & $ \phantom{-}0.291 \pm 0.16 (\rm stat) ^{+0.38} _{-0.44} (\rm syst)$ \\
  \end{tabular} \end{ruledtabular}
\end{table*}
\endgroup

\begin{figure}[t]
    \includegraphics[width=1.0\linewidth]{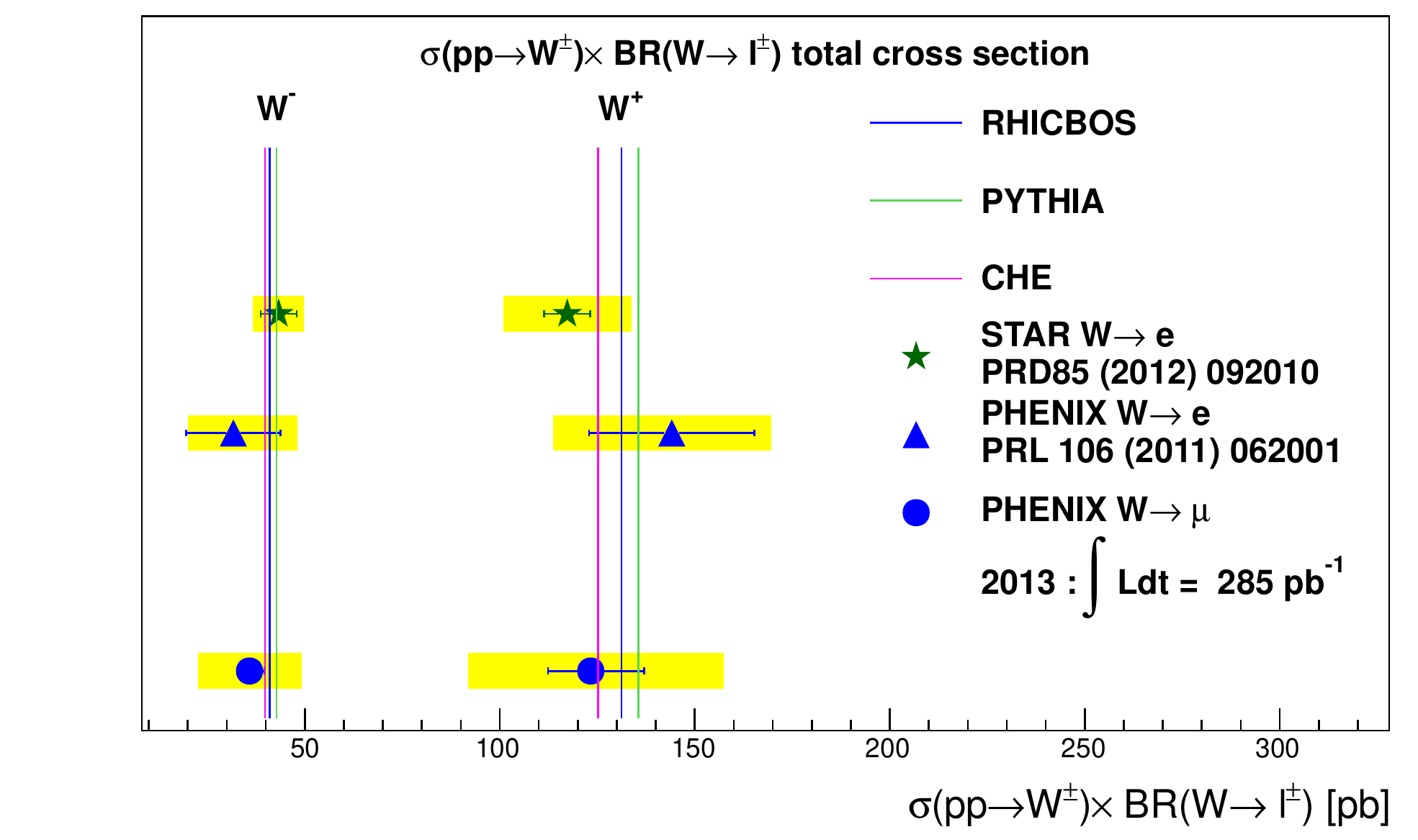}
    \caption{\label{fig:CrossSection} 
The $W^{\pm}\rightarrow\mu^{\pm}\nu$ cross section measured at forward and 
backward rapidity from this measurement averaged over both arms 
$1.1<|\eta|<2.5$ (solid [blue] circles) and central rapidity measurements 
from PHENIX $-0.35<{\eta}<0.35$ (solid [blue] 
triangles)~\cite{Adare:2015gsd} and STAR $-1.0<{\eta}<1.0$ (solid [green] 
stars)\cite{Adamczyk:2014xyw}. The horizontal bars and shading [yellow] 
show the statistical and systematic uncertainties, respectively.  The 
vertical-line estimates from the NLO generators are from left to right: 
{\sc che} [purple], {\sc rhicbos} [blue], and {\sc pythia6.4}, 
using TuneA and a k-factor of 1.4 [green].
}
\end{figure}

Apart from these contributions to the systematic uncertainties, various 
consistency checks were performed to ensure that signals are reliably 
extracted and the single spin asymmetries are correct. The asymmetries 
were tested with randomized helicity patterns to ensure that no false 
asymmetries and no hidden systematic uncertainties were 
present. When changing either the momentum range or the $W_{\rm ness}$ range, 
the amount of background events rapidly grows and the asymmetries all 
become consistent with zero as expected.

\section{Results}

Figure \ref{fig:CrossSection} shows the extracted total cross sections 
for inclusive $W^\pm\rightarrow\mu^\pm$ production in $p$$+$$p$ 
collisions at a center-of-mass energy of 510 GeV.  The cross sections 
are consistent within uncertainties with previous measurements at this 
energy from central $W$$\rightarrow e$ decay channels 
\cite{Adare:2010xa,STAR:2011aa} and with the expected NLO predictions. 
The uncertainties are dominated by the large uncertainty on the 
extracted signal-to-background ratios but are comparable with the 
previously published PHENIX results at central rapidities.

\begin{figure}[t]
\includegraphics[width=1.0\linewidth]{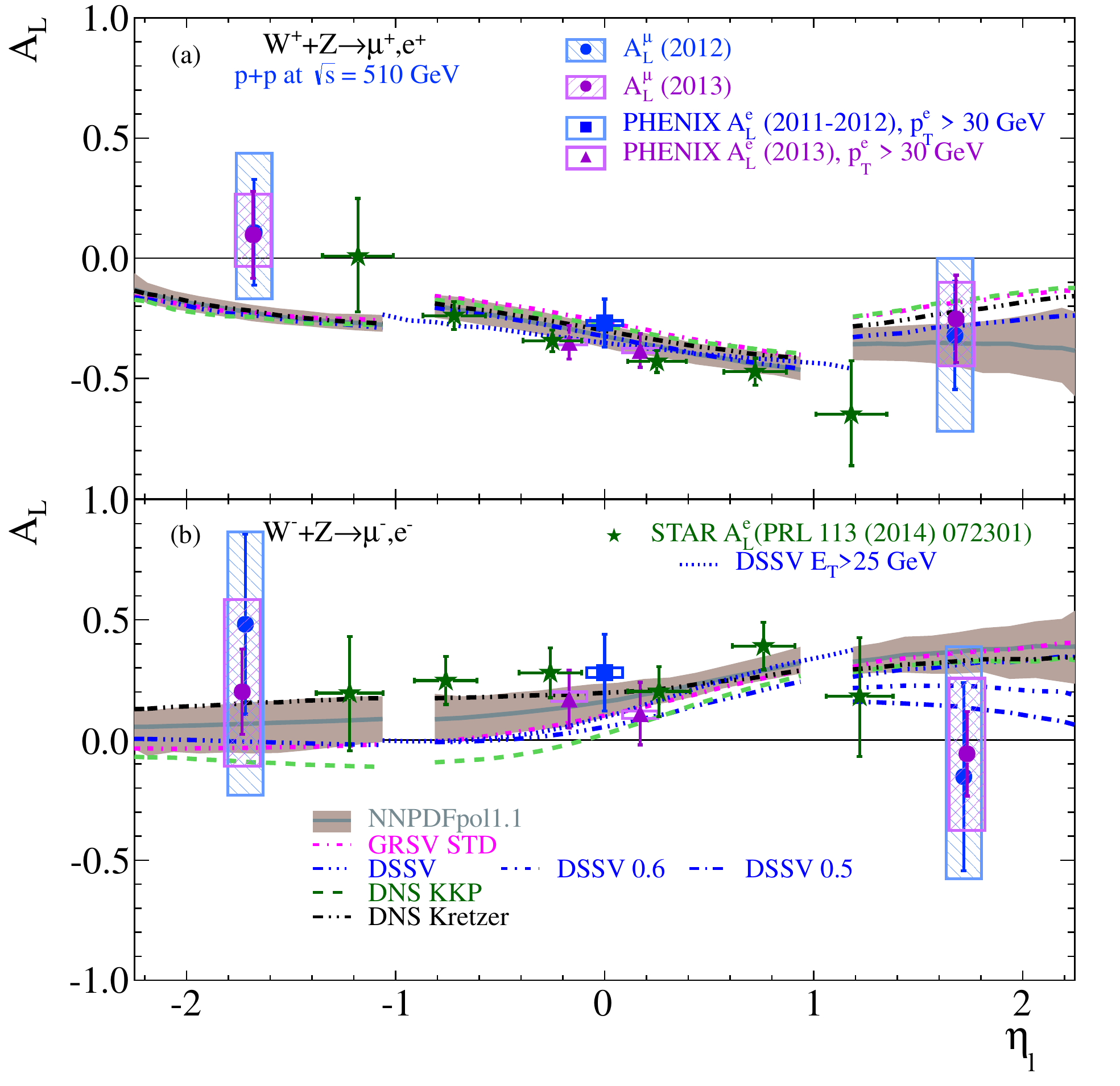}
    \caption{\label{fig:Asymmetry} 
Longitudinal single spin asymmetry, $A_{L}$ for (a) 
$W^{+}+Z{\rightarrow}\mu^{+}, e^{+}$ and (b) 
$W^{-}+Z{\rightarrow}\mu^{-}, e^{-}$.  The PHENIX results are for the 
current combined two muon arms for 2012 (solid [blue] circles) and 2013 
(solid [purple] circles), 
$p^{\mu}_T>16$~GeV , and for previously published 
2010$+$2012, $p^{e}_T>30$~GeV (solid [blue] squares) and 2013, 
$p^{e}_T>30$~GeV (solid [purple] 
triangles)~\protect\cite{Adare:2015gsd}. The STAR results (solid 
[green] stars)~\protect\cite{Adamczyk:2014xyw} are for combined 
2011$+$2012. Also shown are the statistical error bars and systematic 
uncertainty boxes.  The curves depict helicity PDF parameterizations 
from various global fits described in the text that are calculated using 
the polarized NLO generator {\sc che}.
}
\end{figure}


The longitudinal single-spin asymmetries, $A_{L}$, measured at forward 
and backward rapidities are shown in Fig.~\ref{fig:Asymmetry}(a) for 
positive and Fig.~\ref{fig:Asymmetry}(b) for negative $W+Z$ decay muon 
candidates.  The two individual single spin asymmetries from the two 
colliding beams have been combined after correcting for background.  
Vertical lines and boxes show the statistical and systematic 
uncertainties, respectively.  The curves depict parameterizations for the 
quark and anti-quark helicity PDFs based on various global 
fits~\cite{deFlorian:2009vb,Nocera:2014gqa,deFlorian:2005mw,Gluck:2000dy} as 
evaluated for the $W+Z\rightarrow \mu$ process at NLO  
in the strong coupling using the {\sc che} generator~\cite{deFlorian:2010aa}. 


For the NNPDFpol1.1 set, the uncertainty bands based on their 100 replicas 
are also displayed in Fig.~\ref{fig:Asymmetry}.  At forward $\mu^-$ 
rapidities, the DSSV08 curves for two scenarios in which $\Delta 
d(x)/d(x)$ approaches unity when $x$ is approaching unity are also 
displayed for comparison. The previously published central 
$W(+Z)\rightarrow e$ asymmetries from the STAR experiment 
\cite{Adamczyk:2014xyw} and PHENIX \cite{Adare:2015gsd} are also shown. 
2013 results from STAR are still expected \cite{Xu:2017dgx}.

These asymmetries show the first muon single spin asymmetry results from 
$W+Z$ decays at pseudorapidities $|\eta| >1 $ of the decay lepton. They 
help determine the valence and sea quark helicities at different 
momentum fractions than at central rapidities. The uncertainties are 
substantial due to the large systematics on the signal extraction and 
the relatively small signal fractions in the selected data sample. The 
behavior of the asymmetries is generally consistent with the 
parameterizations although the forward $\mu^-$ asymmetry is below the 
DSSV08 curve. While the predicted asymmetries, including a scenario where 
the d-quark polarization changes sign and becomes positive at very large 
$x$ ($x>0.5$), are more compatible with this result, the precision is 
not sufficient to actually confirm it. The backward $\mu^-$ asymmetries 
are at the upper limit of the uncertainty bands, which is similar to the 
central measurements and indicates a $\Delta\bar{u}(x)$ larger than the 
central values obtained in the global fits without the RHIC $W$ 
measurements. The forward $\mu^+$ asymmetries are in agreement with the 
parameterizations, while the backward asymmetries prefer substantially 
smaller asymmetries. Based on the helicity parameterizations, the 
asymmetries are dominated by the well-known up-quark helicities. The 
relatively small NNPDF uncertainty band is dominated by the anti-down 
quark helicity uncertainties. However, for $W^+$ production at forward 
rapidities and our transverse momentum selection, there is always a 
mixture of up and anti-down flavors from either proton that contributes 
at a rather different $x$. It is possible that a higher unpolarized 
anti-down quark component reduces the size of the asymmetries. Such 
uncertainties in the unpolarized PDFs are not included in the 
uncertainty bands of the asymmetry parameterizations. This larger 
unpolarized contribution could explain our surprisingly small backward 
$W^+$ asymmetries.

The total $W$ boson production cross sections for $p$$+$$p$ collisions at
$\sqrt{s}=510$ GeV for 
$\sigma(W^+{\rightarrow}\mu^+)$ and $\sigma(W^-{\rightarrow}\mu^-)$ are
$123.31 ^{+13}_{-11} (\rm stat)^{+34}_{-31} (\rm syst)$ pb and  
$35.80 ^{+3.9}_{-3.0} (\rm stat)^{+13}_{-13}(\rm syst)$ pb,
respectively.   The corresponding single-spin asymmetries, including 
their uncertainties, are summarized in Table~\ref{tble:Systs}.

\section{Summary}

In summary, PHENIX has measured the first longitudinally polarized 
single spin asymmetries in $W\rightarrow \mu$ production at decay lepton 
pseudorapidities larger than unity. The asymmetries from global fits of 
previous longitudinally polarized world data are mostly consistent with 
our results. However, our data also shows a tendency for anti-up quark helicities 
to be closer to the upper limit of the previously extracted 
uncertainties. These measurements will play a major role in reducing the 
uncertainties in future global helicity fits over a larger $x$ range 
than previously covered.


\begin{acknowledgments}  

We thank the staff of the Collider-Accelerator and Physics
Departments at Brookhaven National Laboratory and the staff of
the other PHENIX participating institutions for their vital
contributions.  
We also thank D. DeFlorian and E. Nocera for their support in 
setting up the {\sc che} $W$ simulations with their NNPDFpol 
inclusion and W. Vogelsang for many insightful discussions 
since before the actual measurements were recorded and analyzed. 
We acknowledge support from the
Office of Nuclear Physics in the
Office of Science of the Department of Energy,
the National Science Foundation,
Abilene Christian University Research Council,
Research Foundation of SUNY, and
Dean of the College of Arts and Sciences, Vanderbilt University
(U.S.A),
Ministry of Education, Culture, Sports, Science, and Technology
and the Japan Society for the Promotion of Science (Japan),
Conselho Nacional de Desenvolvimento Cient\'{\i}fico e
Tecnol{\'o}gico and Funda\c c{\~a}o de Amparo {\`a} Pesquisa do
Estado de S{\~a}o Paulo (Brazil),
Natural Science Foundation of China (People's Republic of China),
Croatian Science Foundation and
Ministry of Science and Education (Croatia),
Ministry of Education, Youth and Sports (Czech Republic),
Centre National de la Recherche Scientifique, Commissariat
{\`a} l'{\'E}nergie Atomique, and Institut National de Physique
Nucl{\'e}aire et de Physique des Particules (France),
Bundesministerium f\"ur Bildung und Forschung, Deutscher
Akademischer Austausch Dienst, and Alexander von Humboldt Stiftung (Germany),
J. Bolyai Research Scholarship, EFOP, the New National Excellence
Program ({\'U}NKP), NKFIH, and OTKA (Hungary),
Department of Atomic Energy and Department of Science and Technology (India),
Israel Science Foundation (Israel),
Basic Science Research Program through NRF of the Ministry of Education (Korea),
Physics Department, Lahore University of Management Sciences (Pakistan),
Ministry of Education and Science, Russian Academy of Sciences,
Federal Agency of Atomic Energy (Russia),
VR and Wallenberg Foundation (Sweden),
the U.S. Civilian Research and Development Foundation for the
Independent States of the Former Soviet Union,
the Hungarian American Enterprise Scholarship Fund,
the US-Hungarian Fulbright Foundation,
and the US-Israel Binational Science Foundation.

\end{acknowledgments}  




%
 
\end{document}